\documentclass[11pt,a4paper]{article}

\usepackage{jheppub}

\usepackage{amsmath,hyperref,amssymb}
\usepackage{graphicx,url,color}

\usepackage{caption}
\usepackage{subcaption}

\newcommand{\be}{{\bar{1}}}

\DeclareSymbolFont{AMSa}{U}{msa}{m}{n}
\DeclareSymbolFont{AMSb}{U}{msb}{m}{n}
\DeclareMathSymbol{\fieldR}{\mathalpha}{AMSb}{"52}

\newcommand{\beq}{\begin{equation}}
\newcommand{\eq}{\end{equation}}

\def\be{\begin{equation}}
\def\ee{\end{equation}}
\def\bea{\begin{eqnarray}}

\def\eea{\end{eqnarray}}

\def\lab{\label}
\def\f{\phi}

\def\o{\omega}
\def\le{\left}
\def\ri{\right}

\def\m{\mu}
\def\n{\nu}

\def\6{\partial}

\def\lab{\label}

\def\be{\begin{equation}}
\def\ee{\end{equation}}
\def\bea{\begin{eqnarray}}
\def\eea{\end{eqnarray}}
\def\lab{\label}

\def\le{\left}
\def\ri{\right}
\def\6{\partial}

\title{Holographic equilibration in confining gauge theories under external magnetic fields}
\author[a]{T. Demircik,}
\author[b]{U. G\"ursoy,%
}
\affiliation[a]{Faculty of Engineering and Natural Science, Sabanci University, Tuzla, Istanbul, 34956, Turkey}
\affiliation[b]{Institute for Theoretical Physics and Center for Extreme Matter and Emergent Phenomena,\\
Utrecht University, Leuvenlaan 4, 3584 CE Utrecht, The Netherlands}
\emailAdd{tunad@sabanciuniv.edu}
\emailAdd{u.gursoy@uu.nl}
\abstract{We investigate the effect of external magnetic fields on equilibration in the improved holographic QCD theory in the deconfined phase using the AdS/CFT correspondence. In particular we calculate the quasinormal mode spectra in the corresponding black brane solutions and study their dependence on temperature, momentum and magnetic field, both in the scalar and the shear channels. We find complex patterns in the motion of quasinormal modes on the complex plane, including certain cross overs between the lowest lying modes under varying magnetic field, momentum  and temperature. We also discover a critical value of the magnetic field $B_c$ above which the hydrodynamic approximation breaks down, as the imaginary part of the first excited quasi-normal mode in the shear channel becomes smaller than that of the hydro mode.}
\arxivnumber{16--.----}

\begin{document}
\maketitle

\section{Introduction and summary}
\lab{Intro}

Quantum Chromodynamics exhibits a host of interesting phenomena when put under a strong external magnetic field B. These phenomena range from a complicated dependence of the quark condensate on B, resulting in phenomena such as  magnetic catalysis \cite{Gusynin:1995nb}, inverse magnetic catalysis \cite{Skokov:2009qp}, to the existence of new anomalous transport properties resulting in the chiral magnetic and vortical effects, the chiral magnetic wave \cite{Kharzeev:2007jp, Fukushima:2008xe} and production of electric currents as a result of Faraday and Hall effects \cite{Gursoy:2014aka}. Dependence of finite temperature QCD dynamics on magnetic fields are not only of academic interest, as there is a significant possibility of observing the associated phenomena in the non-central heavy ion collision experiments. It is known that rather large magnetic fields are produced by the spectator ions in such non-central collisions \cite{Kharzeev:2007jp, Skokov:2009qp,Tuchin:2010vs,Voronyuk:2011jd,Deng:2012pc,
Tuchin:2013ie,McLerran:2013hla,Gursoy:2014aka}. Many of these estimations predict presence of these magnetic fields even after the formation of the Quark-Gluon Plasma  in such experiments. Therefore, it is crucial to understand the effects of magnetic fields on the dynamics of the quark-gluon plasma. See \cite{Kharzeev:2012ph} for an extensive review of phenomena associated to presence of magnetic fields. There are also multiple indications that the QGP produced in these experiments is strongly interacting \cite{CasalderreySolana:2011us} preventing a direct study in perturbative QCD. 

Together with the lattice and hydrodynamic simulations, the AdS/CFT correspondence \cite{Maldacena:1997re,Gubser:1998bc,Witten:1998qj} is established as an alternative non-perturbative tool in studying such phenomena in QCD-like {\em confining} gauge theories \cite{Witten:1998zw,Gubser:1999pk,Sakai:2004cn,Erlich:2005qh,Herzog:2006ra, Csaki:2006ji,Gursoy:2007cb}. Recently the AdS/CFT studies have also led to many new insights in regard to non-equilibrium, real-time physics \cite{Chesler:2008hg, Chesler:2010bi, Heller:2011ju, Heller:2012km}. Since such real-time problems are hard to study with lattice methods, the non-perturbative tools provided by the AdS/CFT correspondence  becomes especially important in this respect. Equilibration of small fluctuations around a system at hydrodynamic equilibrium is generically determined by the poles of the associated retarded Green functions on the complex plane. In the dual holographic description they correspond to the quasinormal mode (QNM) spectrum of bulk fluctuations on the corresponding black-brane background. These modes can be obtained by studying fluctuations in the bulk, with infalling boundary conditions on the horizon and vanishing Dirichlet boundary conditions near the asymptotically AdS boundary \cite{Son:2002sd}.  

In this paper, we take the first step to study holographic equilibration in QCD-like confining gauge theories under external magnetic fields. QNMs in the context of holography has been studied extensively, see \cite{Berti:2009kk} for a review. Such  studies in the case of realistic, QCD-like confining theories have only recently started to appear \cite{Gursoy:2013zxa, Buchel:2015saa, Janik:2015waa, Ishii:2015gia, Janik:2016btb, Gursoy:2016ggq}. QNMs in systems with magnetic fields have also been studied in the context of holography before, see for example \cite{Janiszewski:2015ura} but, to our knowledge, the problem has not been investigated in the literature in the backgrounds corresponding to confining gauge theories with magnetic fields. For the gravity background we consider the bottom-up holographic models first considered in \cite{Gursoy:2007cb,Gursoy:2007er,Gursoy:2008za,Gursoy:2009jd}. These models, that are called the {\em improved holographic QCD}  (ihQCD) provide a realistic description of the particle spectrum in the low temperature phase, the confinement-deconfinement transition and the thermodynamic properties of QCD in the high temperature phase. 

The ihQCD model has been extended to include the flavor sector in the large flavor number $N_f$, large color $N_c$ (Veneziano) limit in \cite{Casero:2007ae,Iatrakis:2010jb,Jarvinen:2011qe,Alho:2012mh,Arean:2012mq,Arean:2013tja, Alho:2013hsa,Jarvinen:2015ofa}. We shall deal with such extended models, for the obvious reason that the magnetic field couples directly to the flavor sector, and only indirectly to the glue sector. The model contains two sectors. The glue sector is described by gravity in 5D coupled to a dilaton field that corresponds to the $\mathbb{T}\text{r}F^2$ operator. Running of the dilaton field in the holographic direction mimics the RG running of the 't Hooft coupling constant in energy. Secondly, there is the DBI flavor brane sector describing the dynamics of open strings on the glue background. End points of such  open strings correspond to the quarks and anti-quarks on the dual theory. The lowest lying excitation, the open string tachyon field corresponds to the quark condensate \cite{Casero:2007ae, Gursoy:2007er, Jarvinen:2011qe}.  Finally there is a bulk gauge field on the flavor branes that is chosen such that it  corresponds to the magnetic field on the dual field theory. The background we consider in this paper is, then, an asymptotically AdS black-brane solution with an anisotropy in the direction along the magnetic field that is coupled to the bulk gauge field and the aforementioned tachyon field. When the magnetic field is set to zero, the gravitational theory exhibits a first order Hawking-Page \cite{Hawking:1982dh}  type  phase transition \cite{Gursoy:2008za, Jarvinen:2011qe}. This phase transition corresponds to the confinement-deconfinement transition in the dual confining gauge theory. The  phase transition continues to exist when the magnetic field is turned on, however the value of the transition temperature $T_c$ shifts with $B$ \cite{Rougemont:2015oea, Drwenski:2015sha}.  As we are interested in equilibration in the deconfined phase of the gauge theory, we consider QNMs in the black brane phase, that is for temperatures $T>T_c$\footnote{The black brane solutions in the improved holographic QCD models exist above a certain minimum temperature $T_{min}<T_c$. Therefore, one may also study the QNMs in the range $T_{min}<T<T_c$. These solutions were argued to correspond to a supercooled phase of the theory in \cite{Gursoy:2013zxa}
 and a study of the QNMs for $B=0$ was presented in this paper.}. 

One can think of the spectrum of QNMs in the presence of magnetic fields as analogs of the Landau levels in the deconfined phase of the confining gauge theory. 
We calculate the QNM spectrum on this black-brane by numerically solving the fluctuations that correspond to the {\em shear} and the {\em scalar} modes. We consider the few  lowest lying modes at finite momentum $\vec{k}$, that we take parallel to the magnetic field $\vec{B}$ for simplicity\footnote{Otherwise decomposition of these fields under Lorentz little group becomes complicated.},  and study their dependence on $B$, $k$ and temperature  $T$. Various interesting phenomena arise from our studies: 
\begin{itemize}

\item We observe two distinct classes of QNMs. Firstly there is a class of QNMs that possess both imaginary and real parts. Secondly there is a QNM that is purely imaginary and remains purely imaginary for any value of $B$, $T$ and $k$. This kind of purely imaginary QNMs were observed before in electrically charged AdS-RN black brane solutions, see for example \cite{Maeda:2006by, Janiszewski:2015ura}. However, the latter paper does not report any QNMs of this type in magnetically charged black brane solutions \cite{D'Hoker:2009mm} whereas we find an example of such modes here. Whether this mode is a part of series of infinitely many purely imaginary QNMs, as in \cite{Maeda:2006by} remains to be seen. In our studies we were able to find only one such example. 
 
 \item The absolute values of the real and the imaginary parts of most of the QNMs increase with increasing magnetic field for the scalar and the shear channels and for small momentum values, as also observed in \cite{Janiszewski:2015ura} before. This has an important implication for the equilibration process in the presence of magnetic fields. Since the equilibration time is inversely proportional to the imaginary part of the quasinormal modes, $\tau\sim -1/\textrm{Im}\o$, larger B implies shorter equilibration times. The only exceptions to this pattern are the purely imaginary mode in the scalar channel, figure \ref{fig2}(b) and two modes in the shear channel, figure \ref{fig6}(b). 
 
\item The motion of QNMs with changing momentum exhibit {\em crossing phenomena}. In particular, we observe that the imaginary part of the QNMs both in the shear  and the  scalar channels cross as $k$ increases, see figures \ref{fig7}(b) and  \ref{fig3}(b) respectively. This means that the dominant mode  that controls the equilibration process depends on the value of $k$. In particular, as one observes in figure \ref{fig7}(b), this crossing happens between the hydro mode and the second excited quasi-normal mode in the shear channel, signalling a breakdown of hydrodynamics beyond a certain value of momentum (that depends on T and B). This phenomenon was also observed in \cite{Janik:2015waa, Ishii:2015gia, Janik:2016btb}. 

\item The same crossing phenomenon also appears with changing temperature. We observe this only in the scalar channel, see figure \ref{fig1}(b). This does not signal breakdown of hydro approximation as there exists no hydro mode in the scalar channel. However it does mean that the dominant mode that controls equilibration changes above a certain value of T (that depends on B and momentum k). 

\item Finally, as perhaps the most interesting finding in this paper, we find that the aforementioned crossing phenomenon also occurs under changing magnetic field, see figure \ref{fig6}(b). Here we observe that the hydrodynamic mode in the shear channel is overtaken by the first excited quasi-normal mode above a certain value of B, that depends on the values of T and k. This is {\em breakdown of hydro approximation beyond a critical magnetic field B}.    

\end{itemize} 
The rest of the paper is organized as follows. In the next section we present our 5D gravity set-up and the black brane background in detail. Section \ref{sec::qnm} contains our main studies of the quasinormal mode spectra. In the last section \ref{Discussion} we discuss our findings and their interpretation from the point of view of the boundary gauge theory. Appendix A detail our calculations.     

\paragraph{Note added:} In the first version of this paper we reported an instability that appeared for large values of momentum. This turned out to be due to an error in our code that we fixed in this version and we do not find any perturbative instability in the system.

\section{Gravity setting}

The ihQCD \cite{Gursoy:2007cb,Gursoy:2007er,Gursoy:2008za,Gursoy:2009jd} is a string theory inspired bottom up holographic model that describes the {\em glue} sector of QCD-like confining gauge theories at large-$N_c$ and large 't Hooft coupling. At high temperatures $T>T_c$, the model gives a realistic description of the glue plasma as comparison to lattice studies in \cite{Panero:2009tv} shows. In this paper we are interested in the effects of the magnetic field in the holographic description of the Quark-Gluon Plasma. Since the magnetic field couples the QGP through the  {\em flavor} sector, i.e. the quarks, one first needs to add the flavor sector in the ihQCD theory. This can be achieved by considering $N_f$ pairs of flavor branes and anti-branes embedded in the glue background. Let us denote the gravitational action corresponding to the glue and the flavor sectors by $S_g$ and $S_f$ respectively. The total gravitational action is  
\begin{eqnarray}
S=S_g+S_f\, .
\end{eqnarray}
Since we are after non-negligible effects of the magnetic field on the QGP then the flavor sector should be taken as the same order as the glue. On the large $N_c$ field theory side, this is achieved by considering the Veneziano limit, $N_f\rightarrow\infty$, $N_c\rightarrow\infty$ and $N_f/N_c=x=$finite and $\lambda=g^2N_c=$fixed. For simplicity we consider finite but small $x$, in practice we shall take $x=0.1$. On the gravity side, finite $x$ means taking into account the backreaction of the flavor sector on the glue background. The latter holographic theory was first constructed  in \cite{Casero:2007ae,Iatrakis:2010jb,Jarvinen:2011qe,Alho:2012mh,Arean:2012mq,Arean:2013tja, Alho:2013hsa,Jarvinen:2015ofa} and called the holographic Veneziano QCD model,  V-QCD for short. This is the model we use in this work. 
 
\subsection{The glue sector}
The glue sector is described by the two-derivative Einstein-dilaton action \cite{Gursoy:2007cb,Gursoy:2007er,Gursoy:2008za,Gursoy:2009jd}
\begin{eqnarray}
\lab{gaction}
S_g=M^{3}_p N^{2}_c \int d^5x \sqrt{-g}\left(R-\frac{4}{3}(\partial \phi)^2+V_g(\phi)\right),
 \end{eqnarray}
where $M_p$ is the 5D Planck mass fixed by matching the high temperature limit of non-interacting fermions and bosons \cite{Alho:2012mh}, $(Ml)^3=(1+7x/4)/45\pi^2$, where $l$ is the AdS radius. The metric is dual to the energy momentum tensor and the dilaton corresponds  to the $\mathbb{T}\text{r}F^2$ operator in  the boundary field theory. We further define 
\be\lab{defl}
\lambda=e^\phi\, ,
\ee
for notational convenience. This is identified with the 't Hooft coupling in the ihQCD theory \cite{Gursoy:2007cb}. 

\subsection{The flavor sector}

The flavor action is described by the generalized DBI action \cite{Casero:2007ae,Jarvinen:2011qe} describing the embedding of the flavor branes
\begin{eqnarray}
\lab{faction}
S_f=-\frac{1}{2}M_{p}^3 N_c\mathbb{T}\text{r}\int d^5x\left(V_f(\lambda, T^\dag T)\sqrt{-\det \mathbf{A}_L} +V_f((\lambda, T T^\dag) \sqrt{-\det \mathbf{A}_R}\right),
\end{eqnarray}
where 
\begin{eqnarray}
\mathbf{A}_{L\mu\nu}&=&g_{\mu\nu}+w(\lambda,T)F^{L}_{\mu\nu}+\frac{\kappa(\lambda,T)}{2}\left[(D_\mu T)^{\dag}(D_\nu T)+(D_\nu T)^{\dag}(D_\mu T)\right],\nonumber \\
\mathbf{A}_{R\mu\nu}&=&g_{\mu\nu}+w(\lambda,T)F^{R}_{\mu\nu}+\frac{\kappa(\lambda,T)}{2}\left[(D_\mu T)(D_\nu T)^{\dag}+(D_\nu T)(D_\mu T)^{\dag}\right],
\end{eqnarray}
and the covariant derivative is given by 
\begin{equation}
D_\mu T=\partial_\mu T+iTA^L_\mu -iA^R_\mu T.
\end{equation}
Here, $A_L$ and $A_R$ denote the gauge fields living on the flavor D-branes corresponding to the global flavor symmetry $U(N_f)_L\times U(N_f)_R$ on the dual field theory. $T$ is a complex scalar, the open string tachyon,  which transforms as a bifundamental under this flavor symmetry and corresponds to  the quark mass operator. The vector and the axial combinations are given by 
\begin{eqnarray}
\lab{vecax}
V_\mu=\frac{A^L_\mu +A^R_\mu}{2} \quad,\quad A_\mu=\frac{A^L_\mu -A^R_\mu}{2}.
\end{eqnarray}
Following \cite{Jarvinen:2011qe} we choose the tachyon potential as 
\begin{equation}
V_f(\lambda,TT^{\dag})=V_{f0}(\lambda)e^{-a(\lambda)TT^{\dag}}.
\end{equation}
 We also make a simplifying assumption and take $\kappa(\lambda,T)$ and $w(\lambda,T)$ independent of $T$. The potentials $V_{f0}(\lambda)$, $a(\lambda)$, $\kappa(\lambda)$ and $w(\lambda)$ are constrained by requirements from the low energy QCD phenomenology, such as chiral symmetry breaking and meson spectra \cite{Arean:2013tja}. 

We make a final simplification by choosing a diagonal tachyon field
\begin{equation}
T=\tau(r) \mathbb{I}_{N_f},
\end{equation}
that corresponds to $N_f$ light quarks with the same mass in boundary field theory. Holographically, $\tau(r)$ is dual to the quark mass operator and its non-trivial profile is responsible for the chiral symmetry breaking on the boundary theory. 

As mentioned above, we consider small values of the parameter $x=N_f/N_c$  in this work, for which the presence of the magnetic field is not expected to change the phase structure substantially. Therefore the dominant phase should be deconfined and chirally symmetric, $\tau(r)=0$, at high temperatures \cite{Casero:2007ae,Jarvinen:2011qe}. From now on we set $\tau=0$ in the background solution. For relatively small values of the magnetic field, and for $T = \tau = 0$, the flavor action (\ref{faction}) can be expanded as 
\begin{eqnarray}
\lab{faction1}
S_f&=&-M^3_pN_c\mathbb{T}\text{r}\int dx^5 V_f(\lambda)\sqrt{-g}\sqrt{\det(\delta^{\mu}_{\nu}+w(\lambda)^2g^{\mu\rho}F_{\rho\nu})} \nonumber \\
&=&-M^3_pN_cN_f\int dx^5 V_f(\lambda)\sqrt{-g}\left(1+\frac{w(\lambda)^2}{4}F_{\mu\nu}F^{\mu\nu}\right).
\end{eqnarray}
where $F_{\m\n}$ is the field strength corresponding to $V_\m$ in (\ref{vecax}). 

\subsection{Background at a finite magnetic field and temperature}

Our complete action is given  by (\ref{gaction}) and (\ref{faction1}). The background solution we seek should solve the Einstein, the Maxwell and the dilaton equations
 \begin{eqnarray}
&&R_{\mu\nu}-\frac{1}{2}g_{\mu\nu}R-\left(\frac{4}{3}\partial_\mu \phi \partial_\nu \phi-\frac{2}{3}g_{\mu\nu}(\partial \phi)^2+\frac{1}{2}g_{\mu\nu}V_{eff}\right) \nonumber \\
&&\lab{EE} -x\frac{V_b}{2}\left({F_\mu}^\rho F_{\nu\rho}-\frac{g_{\mu\nu}}{4}F_{\rho\sigma}F^{\rho\sigma}\right)=0\, ,\\
&&\lab{DE}
\frac{1}{\sqrt{-g}}\partial_\mu\left(\sqrt{-g}g^{\mu\nu} \partial_\nu\phi \right)+\frac{3}{8}\frac{\partial V_{eff}}{\partial \phi}-\frac{3x}{32}\frac{\partial V_b}{\partial\phi}F^2=0\, ,\\
&&\lab{ME}
 \partial_{\mu} \left(\sqrt{-g}V_bF^{\mu\nu} \right)=0\, ,
 \end{eqnarray} 
 respectively. In general we also have to solve the tachyon equation of motion for the field $T$ in (\ref{faction}). However, as we mentioned above, in this paper we are interested in the deconfined and chirally symmetric phase of the plasma, in which one can consistently set $\tau=0$. In other words, $\tau=0$ is a particular solution of the coupled system that includes the tachyon field that we consider in this paper.  
 
Here we introduced the following combinations of potentials
\be\lab{efpots}
V_b(\lambda)=V_f(\lambda)w(\lambda)^2, \qquad V_{eff}(\lambda)=V_g(\lambda)-xV_{f0}\, . 
\ee
We first introduce a constant background magnetic field in the $U(1)$ part of the vector field (\ref{vecax})
\begin{eqnarray}
\lab{Vmu}
V_\mu =\left(0,-\frac{x_2 B}{2},\frac{x_1 B}{2},0,0\right)\, .
\end{eqnarray}
Then the Maxwell's equations are automatically satisfied by (\ref{Vmu}). Presence of this constant magnetic field in the $x_3$ direction breaks the SO(3) symmetry 
 to SO(2), that is, rotations on the $x_1x_2$ plane. Accordingly, we choose an ansatz for the metric and the dilaton as
 \begin{eqnarray}
 \lab{met}
 ds^2=e^{2A(r)}\left(-e^{g(r)}dt^2+dx_1^2+dx_2^2+e^{2W(r)}dx_3^2+e^{-g(r)}dr^2\right), \qquad \lambda=\lambda(r)
 \end{eqnarray}
 where $r\in[0,r_h]$. The UV boundary is at $r=0$, where we demand  $AdS_5$ asymtotics ($A\to -\log(r)$, $g\to 0$, $W\to 0$ as $r\to 0$). $r_h$ is the location of the horizon where $g(r_h)=-\infty$.  

With the ansatz (\ref{met}), the Einstein's equations (\ref{EE}) yield three second order equations for functions $A(r)$, $g(r)$, $W(r)$ and one constraint equation \cite{Drwenski:2015sha}. The scalar equation of motion (\ref{DE}) is not independent and can be obtained from the Einstein equations. In practice, it can be replaced by the aforementioned first order constraint equation. All in all, one obtains the following system of differential equations of degree seven \cite{Drwenski:2015sha}
\bea\lab{EB1}  
&&  A'' -A^{'2}+ \frac13 \le(W''+W^{'2}\ri) = - \frac49 (\f')^2 \\
\lab{EB2}
&&  g''+ g'(g' + 3A'+W') =\frac{x B^2 V_b}{e^{2A+g}} \\
\lab{EB3} 
&&(W'e^{3A+W+g})' = \frac{x B^2}{2} V_b\, e^{W+A} \\
\lab{EB4}  
&&  3A'(g'+4A'+2W') +g'W' + e^{-g} V_{eff} -\frac43 (\f')^2 = -\frac{x B^2\,  V_b}{2\, e^{4A+g}} \, .
  \eea  
 For the potential $V_g(\lambda)$, we choose the ihQCD potential \cite{Gursoy:2007cb,Gursoy:2007er,Gursoy:2008za,Gursoy:2009jd}
 \begin{equation}
 V_g(\lambda)=\frac{12}{l^2}\left(1+V_0 \lambda +V_1 \lambda ^{4/3}  \sqrt{\log(1+V_2 \lambda ^{4/3}+V_3 \lambda^2}) \right) ,
 \end{equation}
with coefficients
 \begin{eqnarray}
V_0&=&\frac{8}{9} \beta_0,  \quad V_1=1000V_0 ^{\frac{4}{3}}, \quad V_3=100000 k^2 \nonumber \\
V_2&=&\left(23+36\times\frac{51}{121}\right)^2 \frac{b_0}{6561 V_0^{2} V_1^2} 
 \end{eqnarray}
where $\beta_0=22/(3(4\pi)^2)$,  $b_0=9/8$ and $k=(8/9)\beta _0$. The UV asymptotics (small $\lambda$) of this potential match the perturbative large-$N_c$ $\beta$-function and corresponds to  initial conditions of an RG flow with asymptotic freedom in the dual field theory \cite{Gursoy:2007cb}. In the IR (large $\lambda$) this form of the potential guarantees that the dual field theory is confining with a gapped glueball spectrum \cite{Gursoy:2007er}. 

The flavor sector potentials $V_{f0}(\lambda)$ and $w(\lambda)$ can be determined by comparing their asymptotics with lattice and pertubative results. Following \cite{ Jarvinen:2011qe,Alho:2012mh,Arean:2012mq,Arean:2013tja} we choose, 
\begin{eqnarray}
V_{f0}(\lambda)&=&W_0(1+W_1\lambda +W_2 \lambda^2), \\
w(\lambda)&=&\left(1+\frac{3a_1}{4}\lambda\right)^{-\frac{4}{3}},
\end{eqnarray}
with 
\begin{eqnarray}
a_1&=&\frac{115-16x}{216\pi^2}, \qquad W_0=\frac{3}{11}, \qquad W_1=\frac{24+(11-2x)W_0}{27\pi^2 W_0}, \\ 
W_2&=&\frac{24(857-46x)+(4619-1714x+92x^2)W_0}{46656\pi^4 W_0}. 
\end{eqnarray}
This choice of the flavor potentials are motivated by matching the perturbative anomalous dimension of the quark mass operator in the UV and the requirement of chiral symmetry and the meson spectra  in the IR \cite{Arean:2013tja}.

With this choice for the potentials,  equations (\ref{EB1}-\ref{EB4}) can only be solved numerically. 
This system of equations require 7 integration constants. However, one can easily see that the regularity at the horizon fixes two of these integration constants automatically, one in (\ref{EB2}) and (\ref{EB3}) each. Two constants among the leftover 5 are determined by the near UV requirements $g\to 0$ and $W\to 0$. The remaining 3 integration constants are physical; they correspond to the volume of a unit cell in the dual theory, the temperature T and $\Lambda_{QCD}$. Furthermore, the volume just factors out as a multiplicative factor in all dimensionful parameters and cancels out in dimensionless ratios. Therefore, together with the magnetic field, our solutions will only be characterized by three parameters: $B, T$ and $\Lambda_{QCD}$. These numerical solutions are first constructed in \cite{Drwenski:2015sha} by matching the background solution with $B=0$ near the boundary. Here we obtain the same solutions 
%
%
with different temperature values in the interval $T/T_c \in[1.00208,1.84416]$ and different magnetic field values in the interval $eB_{phys} \in [0.05978,3.34753]GeV^2,$ where $T_c$ is the critical temperature for the decofinement transition for $B=0$ \cite{Gursoy:2008za}, $e$ is the fundamental electric charge and we defined the ``physical" magnetic field $B_{phys}$ in the dual field theory
\begin{eqnarray}
 \lab{EBph}
 B_{phys}=\frac{B}{l^2}, 
\end{eqnarray}
where $l\approx\frac{0.00161482}{247MeV}$  \cite{Drwenski:2015sha}. The black brane solutions that we obtain by solving the equations (\ref{EB1}-\ref{EB4}) are identical to those in \cite{Drwenski:2015sha}. Therefore we shall not show them here. Instead, we directly  focus on the quasinormal mode spectrum in the next section. 

\section{Quasinormal mode spectra}
\lab{sec::qnm} 

Quasinormal modes (QNMs) are linear fluctuations around the background  that satisfy infalling boundary condition at the horizon and vanishing Dirichlet boundary condition at the boundary. Therefore we first need to determine these linear fluctuation equations around the background solution (\ref{met}) of the previous section. It is standard to take fluctuations in a plane wave form times and $r$-dependent function determined from the linearized Einstein and Maxwell equations. Setting the plane wave part is nontrivial for the model we consider here for the following reason. The background metric (\ref{met}) has $SO(2)$ symmetry around the $x_3$-axis, because of the presence of a constant magnetic field along that direction. Any plane wave form other than exp$(ikx_{3}-i \omega t)$ would spoil this symmetry structure and would result in coupling between all  metric and gauge field fluctuations. This general problem is very hard to solve and we leave it for future work. Instead, here we consider the case when the momentum $\vec{k}$ is aligned with the magnetic field $\vec{B}$.  The fluctuations we consider in this paper are then of the form
\begin{eqnarray}
g_{\mu\nu} =  g_{\mu\nu}^{(0)}+ g_{\mu\nu}^{(1)}, \qquad V_{\mu} = V_{\mu}^{(0)}+ V_{\mu}^{(1)},   
\end{eqnarray} 
where
\begin{equation}
g_{\mu\nu}^{(1)}=e^{i(kx_3-\omega t)}h_{\mu\nu}(r)\, ,
 \qquad  V_{\mu}^{(1)}=ie^{i(kx_3-\omega t)}v_{\mu}(r)\, .
\end{equation}
 Here and throughout  the paper, the superscript $(0)$ denotes background quantities and the superscript $(1)$ denotes fluctuations. In this case we can decouple the scalar, vector, and tensor-type fluctuations. After imposing the radial gauge 
\begin{equation}
\lab{gauge}
h_{tr}=h_{x_3r}=h_{rr}=h_{r\alpha}=0, \qquad v_r=0,
\end{equation}
we end up with the classification \cite{Kovtun:2005ev}
\begin{eqnarray}
\lab{spin2}
&&\text{Spin}2 \quad (\text{scalar channel}): \qquad h_{\alpha\beta}-\delta _{\alpha\beta} h/2 \\
\lab{spin1}
&&\text{Spin}1 \quad (\text{shear channel}): \qquad h_{t \alpha},h_{x_3\alpha},v_\alpha \\
\lab{spin0}
&&\text{Spin}0 \quad (\text{sound channel}): \qquad h_{tt},h_{tx_3},h_{x_3x_3},h,v_t,v_{x_3},\phi
\end{eqnarray}
where $\alpha =x_1,x_2$ and $h=\sum_{\alpha}h_{\alpha\alpha}$.

We note that the gauge (\ref{gauge}) does not completely fix the diffeomorphism invariance.  Under infinitesimal diffeomorphisms, $x^{\mu}\rightarrow x^{\mu}+\xi^{\mu}$ (where $\xi_\mu=e^{i(kx_3-\omega  t)}\zeta_\mu(r)$), the metric, gauge field and dilaton fluctuations transform as
\begin{eqnarray}
\lab{gt1}
g_{\mu\nu}^{(1)} &\rightarrow & g_{\mu\nu}^{(1)} -\nabla_{\mu}^{(0)} \xi_{\nu}-\nabla_{\nu}^{(0)} \xi_{\mu},\\
\lab{gt2}
V_{\mu}^{(1)} &\rightarrow & V_{\mu}^{(1)} -g^{(0)\tau \lambda} V_{\tau}^{(0)} \nabla_{\mu}^{(0)} \xi_{\lambda}-g^{(0)\tau \lambda} \xi_{\lambda} \nabla_{\tau}^{(0)} V_{\mu}^{(0)},\\
\lab{gt3}
\phi^{(1)}&\rightarrow & \phi^{(1)}-\xi^{\mu}\nabla_{\mu}^{(0)}\phi^{(0)}.
\end{eqnarray}
In this paper we only consider the QNMs in the shear and  the scalar channels. In the next subsection we  investigate temperature, magnetic field and momentum dependence of the shear channel QNM spectrum. In the subsequent subsection, we perform a similar investigation for the scalar channel.

\subsection{Shear channel}

In this subsection, we consider quasinormal fluctuations in the shear channel (\ref{spin1}) by turning on the fluctuations
\begin{eqnarray}
g^{(1)}_{tx_2}&=&e^{i(kx_3-\omega t)}e^{2A(r)}H_{tx_2}(r),\\
g^{(1)}_{x_2x_3}&=&e^{i(kx_3-\omega t)}e^{2A(r)}H_{x_2x_3}(r),\\
V^{(1)}_{x_1}&=&ie^{i(kx_3-\omega t)}v_{x_1}(r).
\end{eqnarray}
where $H_{tx_2}=h_{x_2}^{t}$, $H_{x_2x_3}=h_{x_2}^{x_3}$. Expanding equations (\ref{EE}) and (\ref{ME}) to linear order, we obtain the following system of equations
\begin{eqnarray}
&&H_{tx_2}''(r)+\left[3A'(r)+W'(r)\right]H_{tx_2}'(r)+\left[\frac{-e^{2W(r)}k^2}{e^{g(r)}}-\frac{4B^2e^{-2A(r)}Z(\phi)}{e^{g(r)}}
\right]H_{tx_2}(r)\nonumber\\
&&-\frac{e^{-2W(r)}k\omega}{e^{g(r)}}H_{x_3x_2}-\frac{4Be^{2A(r)}\omega Z(\phi)}{e^{g(r)}}v_{x_1}(r)=0
\lab{sh1}\, ,\\
\nonumber \\
&&H_{x_3x_2}''(r)+\left[3A'(r)+g'(r)-W'(r)\right]H_{tx_2}'(r)+\left[\frac{\omega^2}{e^{2g(r)}}-\frac{4B^2e^{-2A(r)}Z(\phi)}{e^{g(r)}}
\right]H_{x_3x_2}(r) \nonumber\\
&&+\frac{k\omega}{e^{g(r)}}H_{tx_2}+\frac{4Be^{2A(r)}kZ(\phi)}{e^{g(r)}}v_{x_1}(r)=0\, ,
\lab{sh2}\\
\nonumber \\
&&v_{x_1}''(r)+\left(A'(r)+g'(r)+W'(r)+\frac{Z'(\phi)\phi'(r)}{Z(\phi)}\right)v_{x_1}'(r)+\left(\frac{\omega ^2}{e^{2g(r)}}-\frac{e^{2W(r)}k^2}{e^{g(r)}}\right)v_{x_1}(r)\nonumber \\
\lab{sh3}
&&+\frac{B\omega}{e^{2g(r)}}H_{tx_2}(r)+\frac{Be^{2W(r)}k}{e^{g(r)}}H_{x_3x_2}(r)=0,
\end{eqnarray}
where $Z(\phi)=x V_f (\phi) w^2(\phi)/4$. In addition to these, one obtains a constraint equation
\begin{equation}
\lab{sh4}
\frac{k}{2}H_{x_3x_2}'(r)+\frac{e^{2W(r)}\omega}{2e^{g(r)}}H_{tx_2}'(r)+2Be^{2(W(r)-A(r))}Z(\phi) v_{x_1}'(r)=0\, .
\end{equation}
Since we have three second order equations and a first order constraint equation, this system of differential equations can be reduced to a set of two second order equations. Considering the gauge transformations (\ref{gt1}) and (\ref{gt2}) one determines the gauge invariant combinations as 
\begin{eqnarray}
Z_2(r)&=&k H_{tx_2}(r)+\omega H_{x_3x_2}(r),\\
Z_3(r)&=&v_{x_1}+\frac{B}{2k\omega} [k H_{tx_2}(r)-\omega H_{x_3x_2}(r)]\, .
\end{eqnarray}
Then, from equations (\ref{sh1}), (\ref{sh2}), (\ref{sh3}) and (\ref{sh4}) we obtain two second order gauge invariant equations for $Z_2$ and $Z_3$ 
\begin{eqnarray}
\lab{fluceq3}
Z_2''(r)+C_1Z_2'(r)+C_2Z_2(r)+C_3Z_3'(r)=0\, ,\\
\lab{fluceq4}
Z_3''(r)+D_1Z_3'(r)+D_2Z_3(r)+D_3Z_2'(r)=0\, .
\end{eqnarray}
The coefficients $C_1$, $C_2$, $C_3$, $D_1$, $D_2$ and $D_3$ are lengthy and presented   in Appendix \ref{appA}.
We use the shooting method to determine the QNM spectrum. In this procedure, we first numerically integrate equations  (\ref{fluceq3}) and (\ref{fluceq4})  by specifying infalling boundary conditions at the horizon
\begin{eqnarray}
Z_2(r_h-r)&=&(r_h-r)^{-\frac{i\omega}{4\pi T}}\left[b_0+b_1(r_h-r)+...\right]\\ 
Z_3(r_h-r)&=&(r_h-r)^{-\frac{i\omega}{4\pi T}}\left[c_0+c_1(r_h-r)+...\right]
\end{eqnarray}
where $b_1$, $c_1$ etc. can be expressed in terms of $b_0$ and $c_0$ by considering the near horizon expansion of equations (\ref{fluceq3}) and (\ref{fluceq4}). The coefficients $b_0$ and $ c_0$ are arbitrary. We first set $b_0\rightarrow1$ and $c_0\rightarrow0$, solve equations (\ref{fluceq3}) and (\ref{fluceq4}) numerically and obtain the first set of  linearly independent solutions $\lbrace Z^{(1)}_2,Z^{(1)}_3 \rbrace$. We then repeat the same procedure for  $b_0\rightarrow0$ and $c_0\rightarrow1$ and obtain the second set of  linearly independent solutions $\lbrace Z^{(2)}_2,Z^{(2)}_3 \rbrace$. The Dirichlet boundary condition then means that the determinant of the matrix of maximally independent solutions should vanish on the boundary:
\begin{eqnarray}
\lim_{r\to r_c}\det[H]=\lim_{r\to r_c}[Z_2^{(1)}Z_3^{(2)}-Z_2^{(2)}Z_3^{(1)}]=0,
\end{eqnarray}
where $r_c$ is the UV-cutoff. We then adjust the frequency $\omega_n=\Omega_n+i \Gamma_n$ to determine the solutions which satisfy this requirement. Our results are shown in  figures \ref{fig5}, \ref{fig6} and \ref{fig7} below.  We show our results in terms of dimensionless frequency and momentum
\begin{eqnarray}
\bar{\omega}=\frac{\omega}{2\pi T_c},  \qquad\bar{k}=\frac{k}{2\pi T_c}.
\end{eqnarray}

Figure \ref{fig5} shows the temperature dependence of the three lowest QNMs and two purely imaginary modes for  $T/T_c\in[1.0021,1.8442]$, $eB_{phys}=0.2391GeV^2$, $\bar{k}=1$.  The lowest purely imaginary mode (blue curve on figure \ref{fig5}) corresponds to hydrodynamic mode in the limit of $\bar{\omega}\ll 1$ and  $\bar{k}\ll 1$. As expected, only this mode approaches to real axis when we heat the system up. We also observe that it never develops real part.  For the following three lowest modes, the real parts increases when we heat the system up, on the other hand, the imaginary parts move away from the real axis and the gap between different modes become larger with increasing temperature.  We also observe a second purely imaginary mode.  It never develops a real part and it never becomes dominant for this choice of $k$ and $B$. With increasing temperature, its imaginary part again moves away from the real axis. In particular we do not observe any crossing between the quasi-normal modes as a function of T in the shear channel. 

  \begin{figure}[!h]
        \centering
    \begin{subfigure}[h]{0.48\textwidth}
        \includegraphics[width=\textwidth]{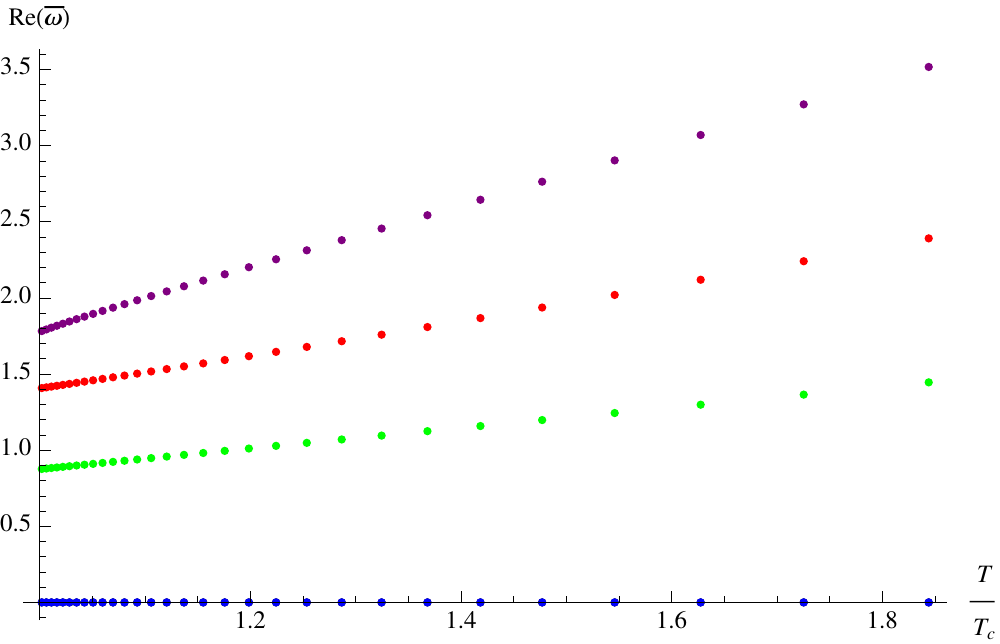}
        \caption{}    
    \end{subfigure}
     ~
    \begin{subfigure}[h]{0.48\textwidth}
        \includegraphics[height=4.5cm]{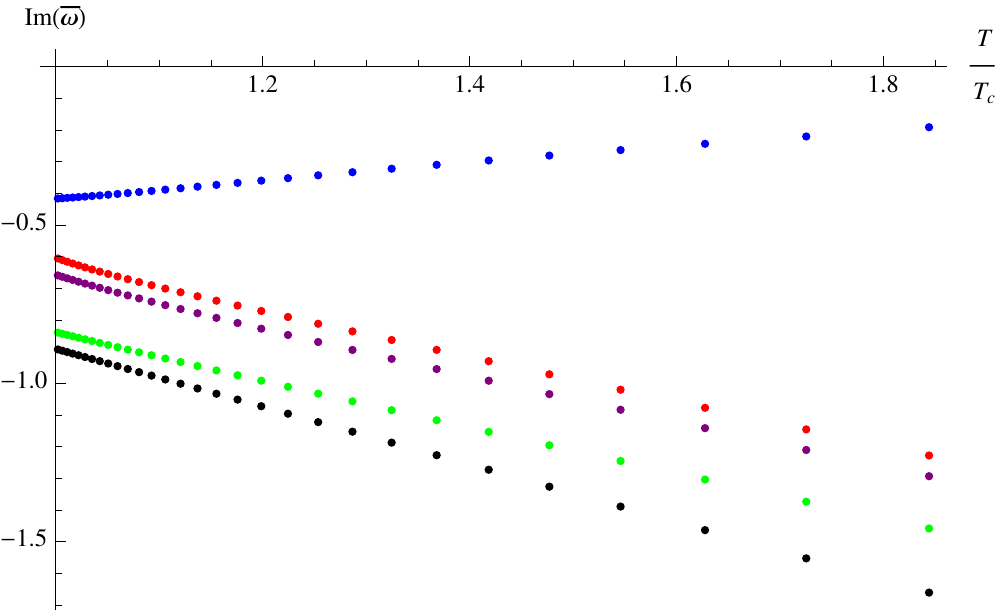}
 \caption{}    
    \end{subfigure}
     ~
    \caption{T-dependence of quasinormal frequencies in the shear channel for temperature values varying between $T/T_c\in[1.0021,1.8442]$ and for fixed $eB_{phys}=0.2391 GeV^2$, $\bar{k}=1$: (a) the real parts and (b) imaginary parts of the three lowest and the two purely imaginary modes. Blue curve is the hydrodynamic mode.}\label{fig5}
\end{figure}
 
In figure \ref{fig6}  we plot the magnetic field dependence of the three lowest QNMs and aforementioned two purely imaginary modes for $eB_{phys}\in[0.1196, 3.3475]GeV^2$, $T/T_c=1.0221$, $\bar{k}=1$. Even though it is not very clear from the plot, we find that their real parts all increase as a function of $B$, except the purely imaginary modes those never develops a real part. The imaginary parts have  more complex behaviour. Modes represented by blue (the hydro mode), purple move downward, red and black move upward and green firstly moves upward then downward with the increasing magnetic field. We observe that behaviour of the imaginary parts depends also on $\bar{k}$ values. For the small $\bar{k}$ all the modes move downward.

Most importantly, we observe a crossing between the blue and red modes for $\bar{k}=1$ and at critical magnetic field value\footnote{This value of $eB_{phys}$ corresponds to a sufficiently small value of  $B$ such that the approximation in (\ref{faction1}) is still valid.} $eB_{phys}=2.9291GeV^2$. This is a crossing between the hydro mode (blue) and the first excited QNM (red), therefore it signals {\em breakdown of the hydrodynamic approximation} above a particular value of the magnetic field $B_c$.

   \begin{figure}[!h]
        \centering
    \begin{subfigure}[h]{0.48\textwidth}
        \includegraphics[width=\textwidth]{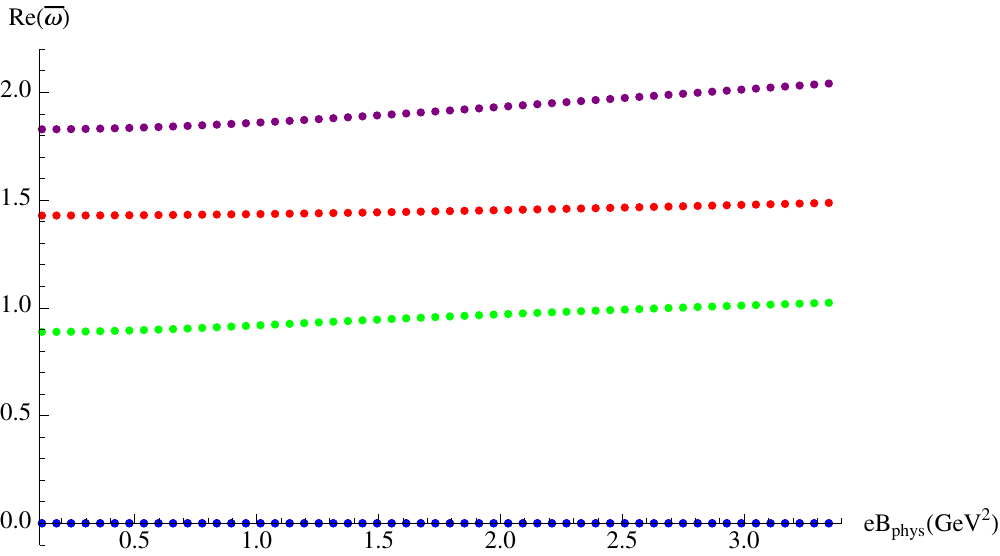}
        \caption{}    
    \end{subfigure}
     ~
    \begin{subfigure}[h]{0.48\textwidth}
         \includegraphics[width=\textwidth]{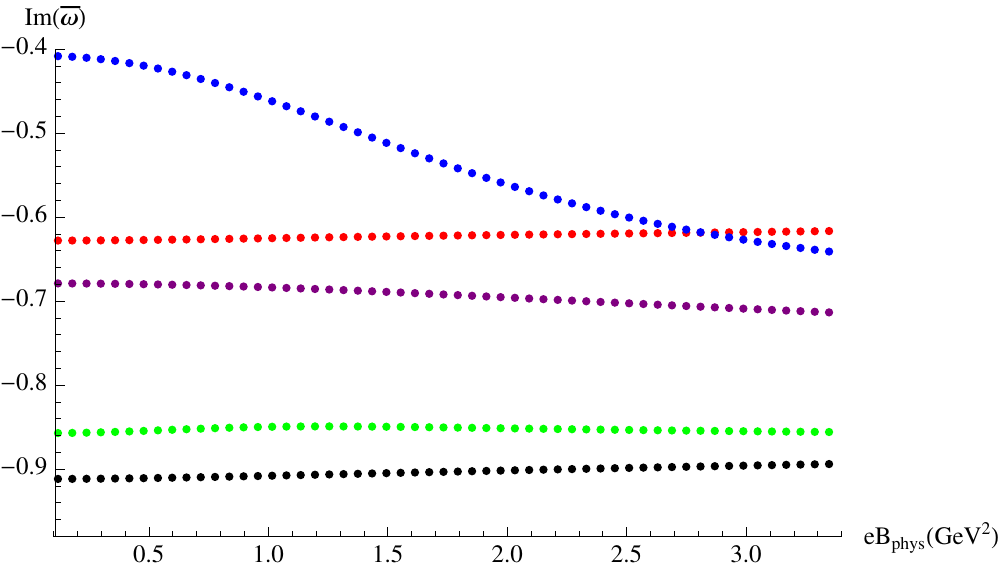}
        \caption{}   
    \end{subfigure}

    \caption{B-dependence of quasinormal frequencies in the shear channel for magnetic field values varying between $eB_{phys}\in[0.1196, 3.3475]GeV^2$ and for fixed $T/T_c=1.0221$, $\bar{k}=1$: (a) the real and (b) the imaginary parts the three lowest and the two purely imaginary mode. Blue curve corresponds to the hydrodynamic mode. We observe that the hydrodynamic approximation breaks down at $eB_{phys}=2.9291GeV^2$.}\label{fig6}
\end{figure}

Finally in figure \ref{fig7}  we look at the momentum dependence of three lowest QNMs and two purely imaginary mode in the shear channel, for $\bar{k}\in[0, 2.2700]$, $T/T_c=1.0221$, $eB_{phys}=0.2391GeV^2$. The real parts increase with the increasing momenta. The momentum dependence of the imaginary parts are more complicated. For every mode we consider in this plot a crossing occurs with another mode at a certain  value of the momentum.  In particular we find that the hydro mode (blue) becomes subdominant with respect to the second excited quasi-normal mode at a critical value $\bar{k}_c=1.3995$. This signals breakdown of the hydrodynamical approximation at large values of momentum. The same phenomenon was also observed in \cite{Janik:2015waa, Ishii:2015gia, Janik:2016btb}. 

  \begin{figure}[!h]
        \centering
    \begin{subfigure}[h]{0.48\textwidth}
        \includegraphics[width=\textwidth]{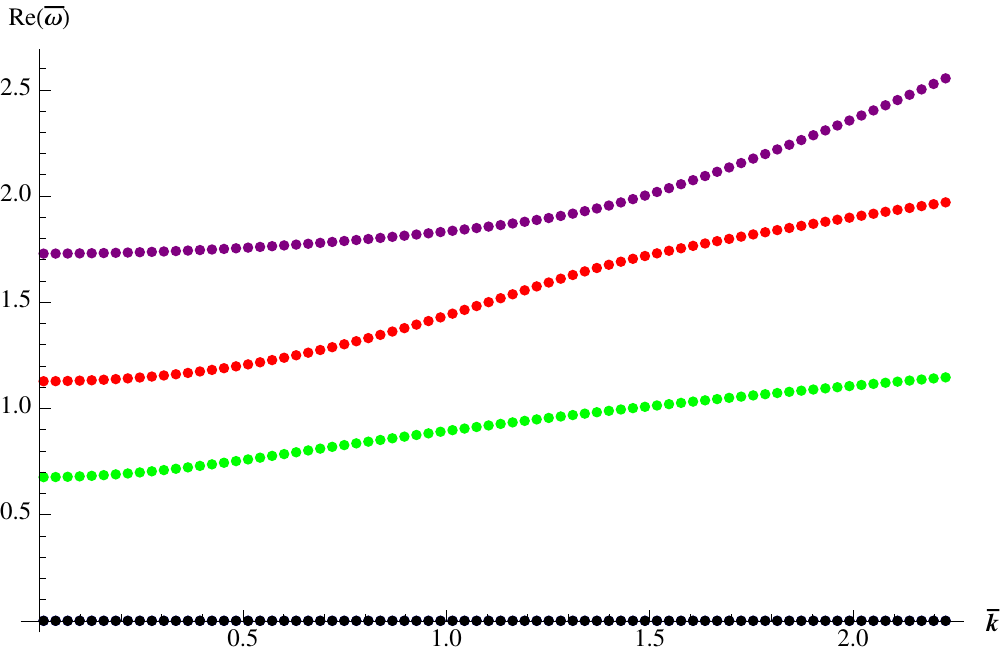}
        \caption{}    
    \end{subfigure}
     ~
    \begin{subfigure}[h]{0.48\textwidth}
         \includegraphics[width=\textwidth]{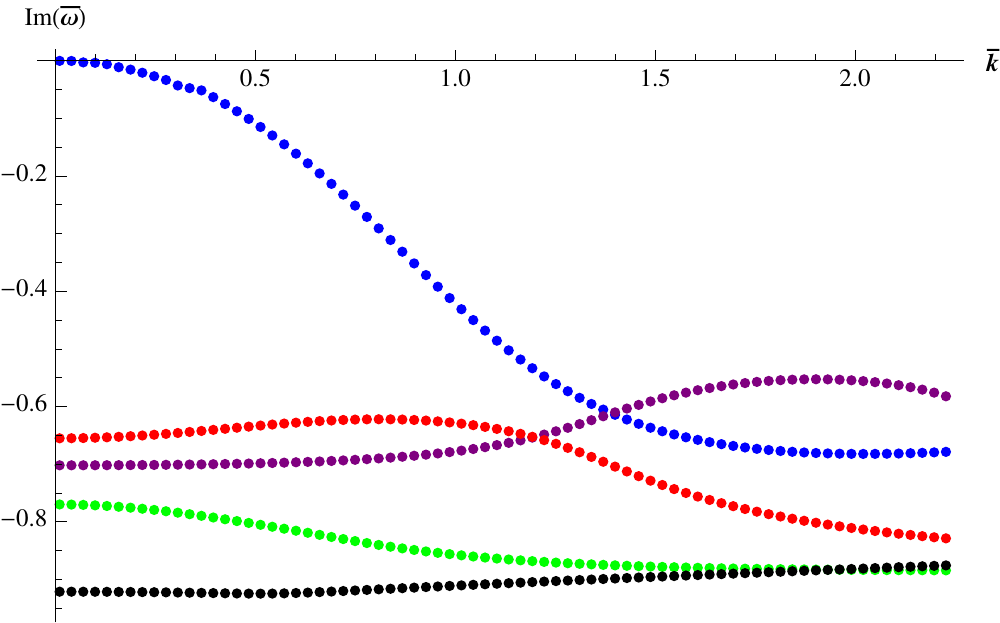}
        \caption{}   
    \end{subfigure}
     
\caption{$\bar{k}$-dependence of quasinormal frequencies in the shear channel for momentum  values varying between $\bar{k}\in[0,2.2700]$ and for fixed $T/T_c=1.0221$and $eB_{phys}=0.2391GeV^2$: (a) the real parts  and (b) the imaginary parts of the three lowest modes and the two purely imaginary modes. Blue curve corresponds to the hydrodynamic mode. We observe that the hydrodynamic approximation breaks down at $\bar{k}_c=1.3995$. }\label{fig7}
\end{figure}

\subsection{Scalar channel}

\begin{figure}[!h]
        \centering
    \begin{subfigure}[h]{0.48\textwidth}
        \includegraphics[width=\textwidth]{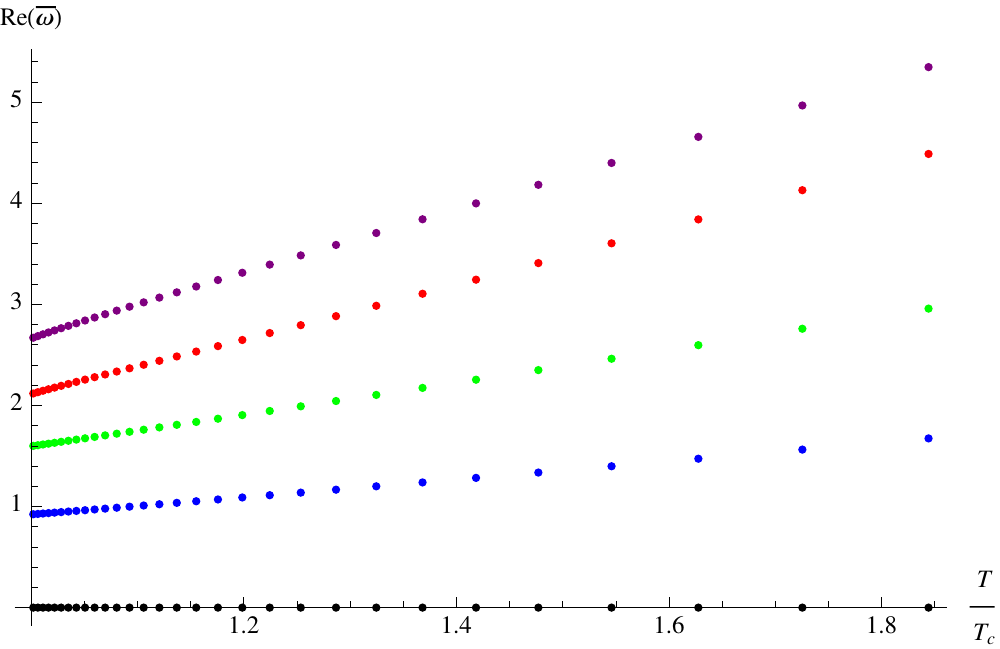}
        \caption{}    
    \end{subfigure}
     ~
    \begin{subfigure}[h]{0.48\textwidth}
        \includegraphics[width=\textwidth]{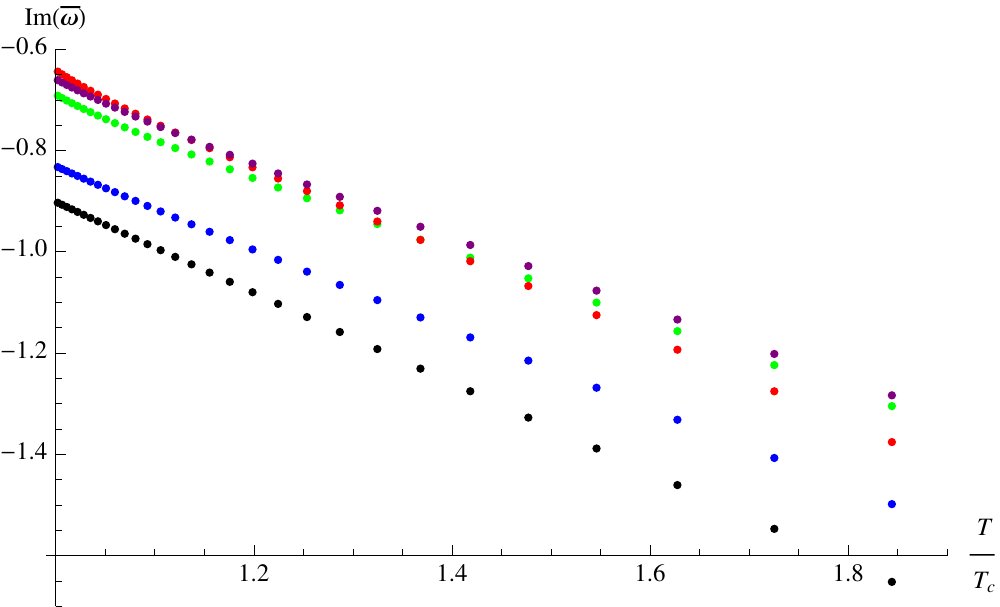}
 \caption{}   
   \end{subfigure}
    \caption{T-dependence of quasinormal frequencies in the scalar channel for temperature values varying between $T/T_c\in[1.0021,1.8442]$ and for fixed $eB_{phys}=0.2391 GeV^2$, $\bar{k}=1$: (a) the real parts and (b) the imaginary parts of the four lowest modes and the purely imaginary mode. We observe crossing between the three lowest lying modes at $T/T_c=1.1371$ and  $T/T_c=1.4185$.}\label{fig1}
\end{figure}

\begin{figure}[!h]
        \centering
    \begin{subfigure}[h]{0.48\textwidth}
        \includegraphics[width=\textwidth]{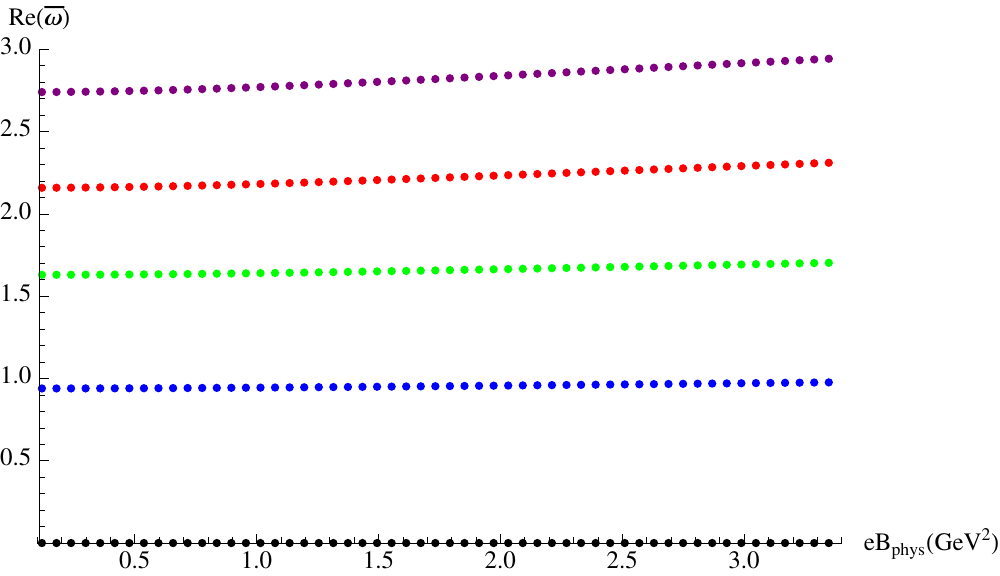}
        \caption{}    
    \end{subfigure}
     ~
    \begin{subfigure}[h]{0.48\textwidth}
         \includegraphics[width=\textwidth]{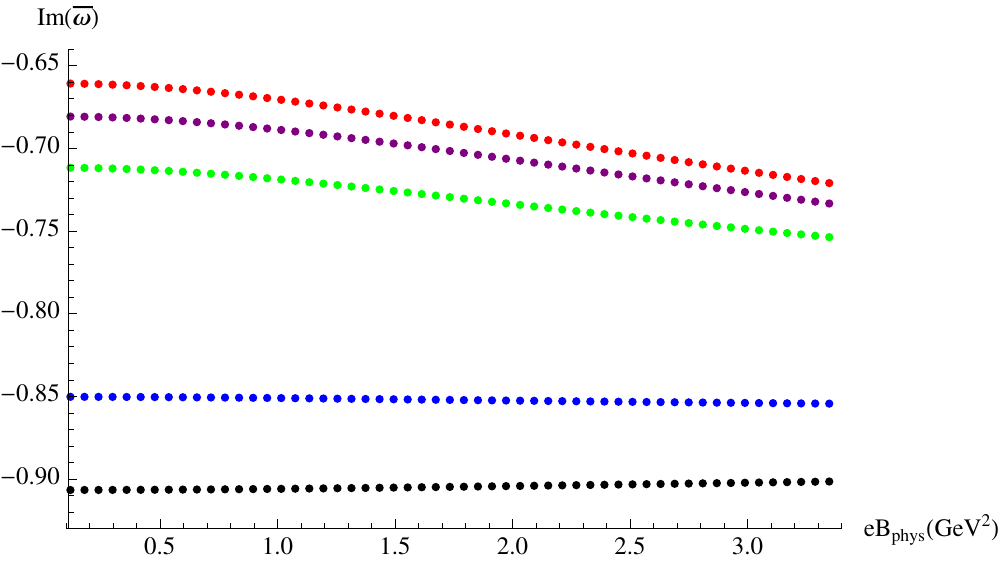}
        \caption{}   
    \end{subfigure}
    \caption{B-dependence of quasinormal frequencies in the scalar channel magnetic field values varying between $eB_{phys}\in[0.1196, 3.3475]GeV^2$ and for fixed $T/T_c=1.0221$, $\bar{k}=1$: (a) the real parts and (b) the imaginary parts the four lowest and the purely imaginary mode.}
       \label{fig2}
\end{figure}

\begin{figure}[!h]
        \centering
    \begin{subfigure}[h]{0.48\textwidth}
        \includegraphics[width=\textwidth]{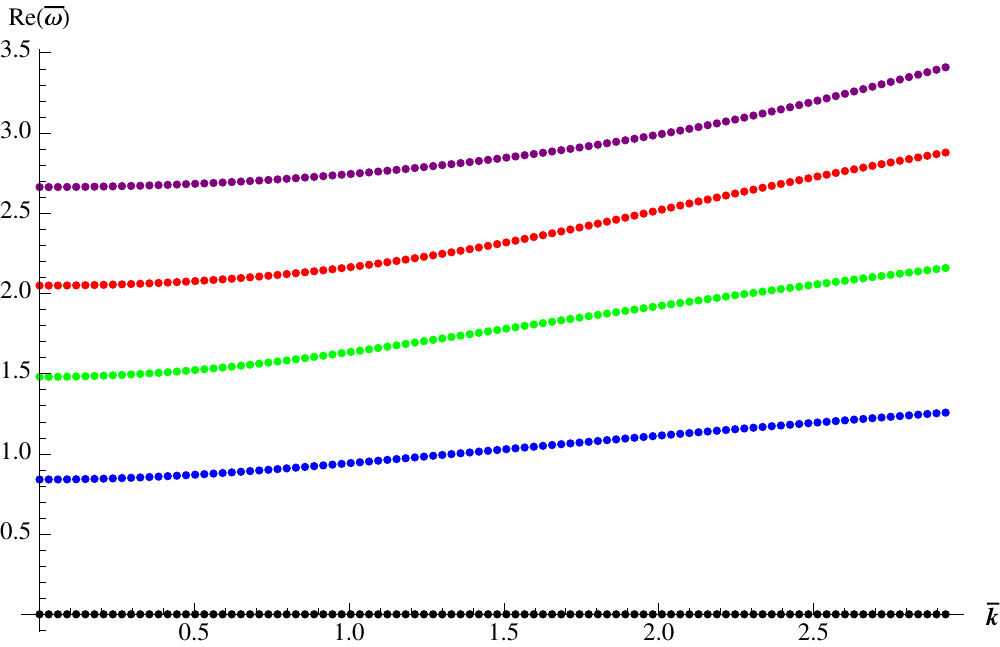}
        \caption{}    
    \end{subfigure}
     ~
    \begin{subfigure}[h]{0.48\textwidth}
         \includegraphics[width=\textwidth]{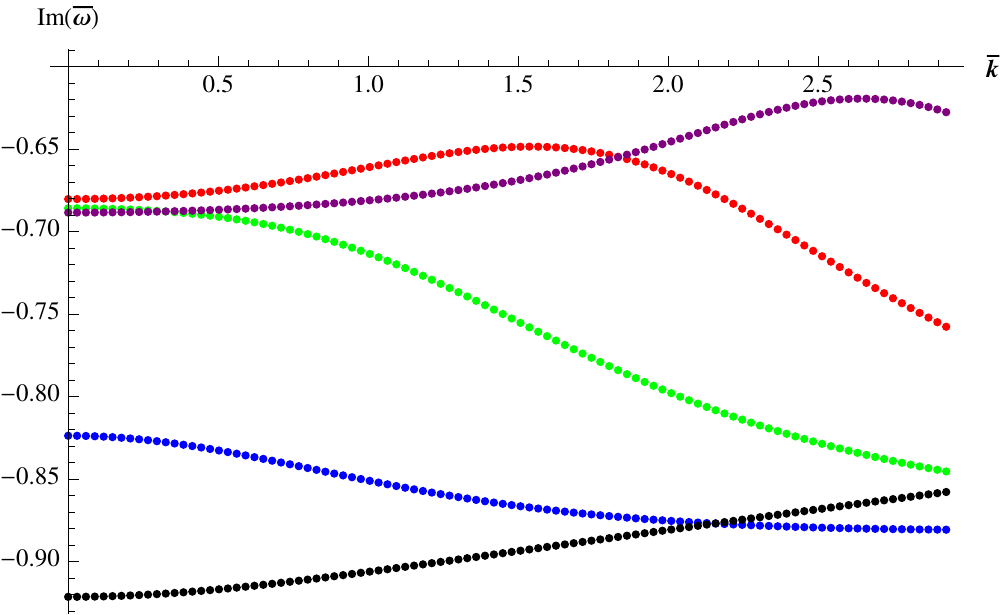}
        \caption{}   
    \end{subfigure}
     \caption{$\bar{k}$-dependence of quasinormal frequencies in the scalar channel for momentum values varying between $\bar{k}\in[0,2.93]$ and for fixed $T/T_c=1.0221$ and $eB_{phys}=0.2391GeV^2$: (a) the real parts and  (b) the imaginary parts of the four lowest modes and   the  purely imaginary mode. We observe various crossings between these modes in (b).}\label{fig3}
\end{figure}

Having observed curious crossing phenomena and breakdown of hydrodynamical approximation as a function of $B$ and $k$ in the shear channel in the previous section, one naturally wonders if the same phenomena occur in the other channels. We answer this question in this section by considering fluctuations in the scalar channel (\ref{spin2}). One important difference from the shear channel is that there is no hydrodynamic mode in the scalar channel. Put differently, there exists no mode whose frequency vanish in the $k\to 0$ limit. Yet, we observe similar crossing phenomena between the different quasi-normal modes as described below. The fluctuation corresponding to the scalar mode is, 
\begin{equation}
g^{(1)}_{x_1x_2}=e^{i(kx_3-\omega t)}e^{2A(r)}H_{x_1x_2}(r)
\end{equation}
where $H_{x_1x_2}=h_{x_2}^{x_1}$ is invariant under the residual gauge transformation (\ref{gt1}) and from now on we denote it by $Z_{1}(r)$. Expanding equation (\ref{EE}) to first order we obtain the gauge invariant fluctuation equation 
\begin{eqnarray}
\lab{sceq}
Z_{1}''(r)+\left[3A'(r)+g'(r)+W'(r)\right]Z_{1}'(r)+\left[\frac{\omega^2}{e^{2g(r)}}-\frac{k^2}{e^{2W(r)+g(r)}}\right]Z_{1}(r)=0.
\end{eqnarray}
To determine the QNM spectrum, we again use the shooting method. We first numerically integrate (\ref{sceq}) by specifying  the infalling boundary condition at the horizon
\begin{equation}
Z_1(r_h-r)=(r_h-r)^{-\frac{i\omega}{4\pi T}}\left[a_0+a_1(r_h-r)+...\right]\, ,  
\end{equation}
where $a_1$ etc. can be expressed as functions of $a_0$  by carrying out the near horizon expansion of (\ref{sceq}). The coefficient $a_0$ on the other hand is arbitrary and we set it to 1. The solution should also satisfy the Dirichlet boundary condition $Z_1(r_{c})=0$ on the boundary.   We then adjust the frequency $\omega_n=\Omega_n+i\Gamma_n$ to determine the solutions that satisfy this requirement. Our results are  shown in figures \ref{fig1}, \ref{fig2} and \ref{fig3}.

In figure \ref{fig1} we consider the temperature dependence of the four lowest QNMs and of the purely imaginary mode for  $T/T_c\in[1.0021,1.8442]$, $eB_{phys}=0.2391GeV^2$, $\bar{k}=1$. We observe that the real parts increase when we heat the system up. The imaginary parts move away from the real axis and the gaps between the different modes become bigger with the increasing temperature as expected. We also observe a purely imaginary mode down on the imaginary axis displayed in figure \ref{fig1}(b).  It never develops a real part and it never becomes dominant at this small value of $\bar{k}$. With increasing temperature, its imaginary part again moves away from the real axis. More importantly we observe a crossing in between the imaginary parts of the three lowest modes. For purple and red modes crossing occurs around $T/T_c=1.1371$ and for green and red modes it occurs around $T/T_c=1.4185$. Even though this does not mean a breakdown of hydro in this case, it is still interesting as it means that different quasi-normal modes control the equilibration of linear fluctuations in the plasma for temperatures above or below these critical values for the temperature. 

Figure \ref{fig2} exhibits the magnetic field dependence of the lowest four QNMs and the aforementioned purely imaginary mode for  $eB_{phys}\in[0.1196, 3.3475]GeV^2$,$T/T_c=1.0221$, $\bar{k}=1$. With increasing magnetic field we observe that the real parts increase (it is not clear from the plot because of the fact that real parts of the QNM frequencies are widely separated) and the imaginary parts move away from the real  axis. We also observe a purely imaginary mode never develops a real part and imaginary part of it moves upward with the increasing magnetic field. We do not find any crossing as a function of B in the scalar channel, at least in the range of B we consider here.  

Finally, the momentum dependence  of the four lowest QNMs and the purely imaginary mode is shown in figure \ref{fig3}  for $\bar{k}\in[0,2.93]$, $T/T_c=1.0221$, $eB_{phys}=0.2391GeV^2$. We find that the real parts increase with increasing momentum. On the other hand, the momentum dependence of the imaginary parts are more complicated. Crossings occur for all the modes shown here at  different values of momentum. 

\section{Discussion}
\lab{Discussion}

In this paper we investigate equilibration of linear fluctuations in the deconfined, and chirally symmetric quark-gluon plasma phase of a strongly coupled large-$N_c$ 4D gauge theory that is confining at low temperatures.  The system is placed under an external magnetic field in the $x_3$ direction and we determine the location of the poles in the retarded Green's function of the stress tensor operator $T_{x_1x_2}$ (the scalar channel) and a mixture of the stress tensor components $T_{t\alpha}$, $T_{x_3\alpha}$ and the current operator $J_{\beta}$ (the shear channel), where $\alpha = x_1, x_2$ and $\beta = x_2,x_1$ respectively.  These poles (especially the lowest modes) control the equilibration process; in particular the equilibration time is well approximated by the imaginary part of the lowest mode $\tau \propto -1/\textrm{Im}\o$. It is therefore quite interesting to determine how this spectrum depends on the parameters $T$, $B$ and the momentum $k$. We carry out this analysis using the holographic dual formulation where these poles are mapped onto the quasinormal modes of the corresponding black-brane solution which we determine by numerically solving the system of linear fluctuation equations using the shooting method.   Our findings are summarized in the list in section \ref{Intro}. Here we discuss our findings in more detail and also provide an outlook.  

\paragraph{B dependence of the QNMs:} The most interesting finding is the breakdown of the hydrodynamical approximation above a certain value of the magnetic field in the shear channel. We observe that the first excited quasi-normal mode becomes more dominant than the hydrodynamic mode, as its imaginary part becomes less negative above a certain value of the magnetic field. As shown in figure \ref{fig6}(b) this happens at $eB_{phys}=2.9291GeV^2$ for  $T/T_c=1.0221$ and $\bar{k}=1$. We note that this finding does not have any direct relevance to heavy ion collisions. As shown in \cite{Gursoy:2014aka}, if we take the proper time value $\tau = 0.5$ fm as the thermalization time, the value of the magnetic field for an off-central collision with a typical impact parameter $b=7$ fm at LHC is around $eB = 0.04 GeV^2$ that is an order of magnitude smaller than the critical value we obtain here. Still, it implies that the hydro approximation is not to be trusted for arbitrarily large values of the magnetic field, consistent with general considerations. We do not observe any such crossing as a function of B in the scalar channel. 

We also find that both in the scalar and the shear channels, the real parts $\textrm{Re}\,\omega$ of all and the imaginary parts $\textrm{Im}\,\omega$ of most of the modes increase with increasing $B$ with the only exception of the purely imaginary mode in the scalar channel and two modes in the shear channel. This means that the corresponding linear fluctuations oscillate faster and become more damped at larger values of $B$. The latter means that the equilibration times $\tau$ become shorter. On the other hand, this should be related to the the thermalization times of the quark-gluon matter produced in heavy ion collisions at RHIC and LHC. One can estimate the thermalization times in the absence(presence) of $B$ in the central(off-central) collisions by matching to the hydrodynamical simulations. One does not seem to find a substantial difference in the thermalization time of the plasma in the presence and absence of $B$. However, as also discussed in \cite{Janiszewski:2015ura}, this does not necessarily imply a contradiction with our results. This is because thermalization and equilibration are two different processes in general. Thermalization refers to a fully non-linear time-dependent evolution and equilibration---that we study in this paper---refers to ring down of {\em linear} fluctuations. Thermalization in heavy ion collisions is expected to be a wild, fully non-linear process that can also be studied holographically \cite{Chesler:2008hg}. One of the most urgent open problems is then to work out the holographic thermalization of the plasma in the presence of magnetic fields. 
As to why the purely imaginary mode in the scalar channel behaves differently and approaches to the real axis with increasing $B$ remains to be understood.  

An interesting open question is to determine the dependence of the linear mode spectra on $B$ in the {\em confined} phase. This would directly provide the Landau levels in the gauge theory, and how they shift with $B$ at strong coupling.

\paragraph{Various crossing phenomena:} 
We find a complicated pattern of motion for the quasi-normal modes, and in particular various crossing phenomena, also as a function of T an k. We find that all the modes we considered in this paper, in both channels, become subdominant for large values of momentum with respect to higher order excitations. This can be seen in figures \ref{fig7}(b) and  \ref{fig3}(b) for the shear and the scalar channels respectively. 
In particular the crossing in figure \ref{fig7}(b) between the hydro mode and the first excited mode signals breakdown of hydrodynamics at large k. This happens at $\bar{k}_c=1.3995$ for $T/T_c=1.0221$, $eB_{phys}=0.2391GeV^2$ in our model. 

Crossings happen less frequently as a function of T. In particular we do not find any crossing in the shear channel, and find only two such cases among the three lowest lying modes in the scalar channel, as shown in figure \ref{fig1}(b). 

\paragraph{Purely imaginary mode:} 
We observe two separate classes of QNMs, one class with both real and imaginary parts and a second class that is {\em purely imaginary}. We find two purely imaginary modes in the shear channel and only one such mode in the scalar channel. 

The first purely imaginary mode in the shear channel has a simple interpretation as it corresponds to the hydrodynamical dispersion relation 
$$ \omega = -i \frac{\eta}{s T} k^2 =  -i \frac{1}{4\pi T} k^2$$
where $\eta$ is the shear viscosity and $s$ the entropy density. The second equation above is confirmed in our backgrounds numerically. 

The other purely imaginary modes are more curious. Such  modes were also observed in charged black-brane solutions, first in \cite{Maeda:2006by} and further discussed in \cite{Janiszewski:2015ura}. The latter paper conjectures an interpretation of such modes, proposing that these poles combine to form a branch cut on the complex frequency plane in the corresponding retarded Green's function in the limit they approach each other. This is a plausible interpretation, however, it does not seem to apply to our case since we were able to locate only one example of such purely imaginary mode, not an entire family as in  \cite{Maeda:2006by}  and \cite{Janiszewski:2015ura}. Whether one can locate the entire family, and whether there exist a limit where they approach each other and form a branch-cut remains to be seen. Another difference between the aforementioned works and our case is that these purely imaginary modes only appear in the electrically charged black-brane solutions in their case, whereas we obtain such an example in the magnetically charged black brane. Physically, it corresponds to an overdamped ring down of the linear fluctuation since it does not possess any real part which would correspond to oscillations in time. These particular fluctuations were associated with backscattering of gravitational waves off the gravitational potential in \cite{Berti:2009kk}. It is plausible that the same interpretation holds in our case. Our purely imaginary modes both in the scalar and the shear channels move up on the imaginary axis with increasing momentum. 

\paragraph{Approximations and a look ahead:} 

Finally let us discuss the various approximations we made in this work. We made three important assumptions in this paper. First, we assumed a finite but small ratio of the number of flavors and color, $x=0.1$ in the Veneziano limit. Even though this is not very realistic for the actual QGP produced at LHC and RHIC, it simplifies our study considerably. In particular we can safely assume that the phase structure of the theory does not diverge significantly from the $x\to 0$ case where there exists only two phases: the confined, chirally broken phase and a deconfined chirally symmetric phase. An intermediate phase where color charge deconfines, yet with a nontrivial quark condensate, as observed in \cite{Alho:2012mh} for example, does not exist in this limit. This assumption allowed us set the tachyon background $\tau=0$. This simplifies the technical analysis considerably, however it raises the obvious questions how does the dependence of quasi-normal modes on B, T and k would be for $x\sim 1$ and/or in an intermediate phase, such as a black-brane with non-trivial tachyon. These are important questions to be addressed in future work. 

Second, we only considered fluctuations with momenta aligned with the magnetic field. This assumption also simplifies the analysis substantially. This is because when k and B are not aligned the problem has less symmetry. Therefore, on one hand classification of modes become harder, on the other hand  one expects more modes couple each other making the coupled system of differential equations more complicated. Yet, this is a restriction that should be overcome in future work. 

Finally we considered a limit where the square-root of the DBI action in (\ref{faction}) can be replaced by the $F^2$ term. This also simplifies the analysis technically but does not correspond to a wide range of parameters. One wonders for example if one would still find breakdown of the hydrodynamic approximation for large B if one lifts this assumption. 

Another obvious omission in our story is the other quasi-normal mode channels. The sound channel is particularly interesting as one would like to know dependence of the dispersion relation of sound mode on the magnetic field. The coupled system of equations in this channel turns out to be too complicated however, therefore we chose to omit this channel altogether in this work.

\section*{Acknowledgements}

We thank Tara Drwenski, Ioannis Iatrakis, Aron Jansen, Cihan Saclioglu and Phil Szepietowski for useful discussions. T. D. thanks ITP, Utrecht university for hospitality where most of this work was completed. This work was supported, in part 
by  the Netherlands Organisation for Scientific Research
(NWO) under VIDI grant 680-47-518, and the
Delta-Institute for Theoretical Physics (D-ITP) that is
funded by the Dutch Ministry of Education, Culture and
Science (OCW). T.D. was supported by the Scientific and Technological Research Council of Turkey (TUBITAK) under the programme 2214-A.

\begin{appendix}
\section{Coefficients in the gauge invariant variable equations of the shear channel}
\lab{appA}
Explicit forms of the coefficients in the gauge invariant equations \ref{fluceq3} and \ref{fluceq4} are 
\begin{eqnarray}
C_1&=&\frac{e^{2W}(2B^2fx(V_b/4)[6fA'+f']-e^{2A}\omega^2[f'+f\lbrace 3A'-W'\rbrace])e^{2A}k^2f^2(3A'+W')}{f(-e^{2(A-W)}\omega^2+f[e^{2A}k^2+4B^2e^{2W}x(V_b/4)])},\\
C_2&=&\frac{\omega^2-e^{2W}k^2f-4B^2e^{-2A}fx(V_b/4)}{f^2},\\
C_3&=&-\frac{4Be^{2W}k\omega x(V_b/4)[f'-2fW']}{-e^{2(A+W)}\omega^2+f[e^{2A}k^2+4B^2e^{2W}x(V_b/4)]},\\
D_1&=&\frac{e^{2A}k^2f(x(V_b/4)[f'+f\lbrace A'+W' \rbrace]+fx(V'_b/4)\phi'+e^{2W}[2B^2fx^2(V_b/4)^2\lbrace 6fA'+f'\rbrace])}{fx(V_b/4)[-e^{2(A+W)}\omega^2+f(e^{2A}k^2+4B^2e^{2W}x(V_b/4))]}\nonumber \\
&&+\frac{e^{2A}k^2f(-e^{2A}\omega^2[x(V_b/4)\lbrace f'+f(A'+W')\rbrace +f x(V'_b/4)\phi'])}{fx(V_b/4)[-e^{2(A+W)}\omega^2+f(e^{2A}k^2+4B^2e^{2W}x(V_b/4))]}, \\
D_2&=&\frac{\omega^2-e^{-2W}k^2f-4B^2e^{-2A}fx(V_b/4)}{f^2},\\
D_3&=&\frac{-2B^3e^{2W}x^2(V_b/4)^2[f'-2fW']}{2k\omega x(V_b/4)[-e^{2(A+W)}\omega^2+f\lbrace e^{2A}k^2+4B^2e^{2W}x(V_b/4)\rbrace]} \nonumber \\
&&+\frac{Be^{2A}[x(V_b/4)(k^2\lbrace 2fA'-f'\rbrace +2e^{2W}\omega^2 \lbrace A'-W'\rbrace)-x(V'_b/4)\phi'(e^{2W}\omega^2 +k^2 f)]}{2k\omega x(V_b/4)[-e^{2(A+W)}\omega^2+f\lbrace e^{2A}k^2+4B^2e^{2W}x(V_b/4)\rbrace]},
\end{eqnarray}
where $f=\log g$, $\phi=\log \lambda$ and the derivatives of $V_b$ are with respect to $\phi$.
\end{appendix}
 
 \bibliographystyle{JHEP}
\bibliography{QNMB-JHEP}

\end{document}